\documentclass[aps,preprint]{revtex4}
\usepackage{graphicx}
\usepackage{float}
\newcommand{\D}{\displaystyle}

\begin{document}
\title {Alpha-decay chains of $^{288}_{173}115$ and $^{287}_{172}115$ in the Relativistic Mean Field theory}

\author{
L.S. Geng$^{1,2,}$\footnote{E-mail address:
lsgeng0@rcnp.osaka-u.ac.jp}, H. Toki$^{1,}$\footnote{E-mail
address: toki@rcnp.osaka-u.ac.jp}, and J.
Meng$^{2,}$\footnote{E-mail address: mengj@pku.edu.cn} }

\affiliation{
$^1$Research Center for Nuclear Physics (RCNP), Osaka University,
Ibaraki, Osaka 567-0047, Japan
\\
$^2$School of Physics, Peking University, Beijing 100871, P.R.
China }

\vspace {30mm}


\begin{abstract}
In the recent experiments designed to synthesize the element 115
in the $^{243}$Am+$^{48}$Ca reaction at Dubna in Russia, three
similar decay chains consisting of five consecutive
$\alpha$-decays, and another different decay chain of four
consecutive $\alpha$-decays are detected, and the decay properties
of these synthesized nuclei are claimed to be consistent with
consecutive $\alpha$-decays originating from the parent isotopes
of the new element 115, $^{288}115$ and $^{287}115$,
respectively\cite{ogan.03}. Here in the present work, the recently
developed deformed RMF+BCS method with a density-independent
delta-function interaction in the pairing channel is applied to
the analysis of these newly synthesized superheavy nuclei
$^{288}115$, $^{287}115$, and their $\alpha$-decay daughter
nuclei. The calculated $\alpha$-decay energies and half-lives
agree well with the experimental values and with those of the
macroscopic-microscopic FRDM+FY and YPE+WS models. In the mean
field Lagrangian, the TMA parameter set is used. Particular
emphasis is paid on the influence to both the ground-state
properties and energy surfaces introduced by different treatments
of pairing. Two different effective interactions in the
particle-particle channel, i.e., the constant pairing and the
density-independent delta-function interaction, together with the
blocking effect are discussed in detail.

\end{abstract}
\maketitle

\section {Introduction}
Since the prediction of the existence of superheavy islands in
1960s \cite{nilsson.69, mosel.69}, the synthesis of superheavy
elements has been a hot topic in nuclear physics. Following
numerous ground breaking technical developments \cite{hof.98} in
the last three decades, the process of synthesizing superheavy
elements has been sped up dramatically. From 1995 to 1996, Hofmann
et al. \cite{hof.951,hof.952,hof.96,hof.98} at GSI in Germany
successfully produced the elements  Z=110, 111, and 112  by using
low-energy heavy-ion collisions. In January 1999, the new element
Z=114 was reported at Dubna in Russia \cite{ogan.991, ogan.992}.
Two years later, the element Z=116 was also reported at Dubna
\cite{ogan.01}. In August 2003, in the reaction
$^{243}$Am+$^{48}$Ca held at Dubna \cite{ogan.03}, with a beam
dose of $4.3\times 10^{18}$ 248-MeV and 253-MeV $^{48}$Ca
projectiles, nine new odd-Z nuclei originating from the isotopes
of the new element 115, $^{288}115$ and $^{287}115$, were
produced, respectively. So far, all elements with $110\le Z\le
116$ have been produced successfully in laboratory. All these
exciting discoveries have greatly extended our knowledge about
superheavy nuclei around the predicted superheavy islands and
drawn more and more attention from the theoretical side.

The experimental progress has led to a large-scale investigation
of superheavy nuclei by both refined macroscopic-microscopic (MM)
models such as the finite-range droplet model with folded-Yukawa
single-particle potentials (FRDM+FY) \cite{moller.97} or the
Yukawa-plus-exponential model with Woods-Saxon single-particle
potentials (YPE+WS) \cite{munt.03}, and microscopic models such as
the Skyme-Hartree-Fock-Bogoliubov method \cite{cwiok.99} and the
latest relativistic mean field model\cite{bend.99,
lala.96,meng.00,long.02,ren.01}. Exploration for the next
so-called "superheavy element island", i.e, the next spherical
doubly magic nucleus, is a dream for physicists for the past
several decades. There are already many works in this respect (see
Ref. \cite{cwiok.99, bend.99, lala.96} and references therein).
Possible candidates predicted by different theories  are
$^{298}_{184}$114, $^{292}_{172}$120 or even $^{310}_{184}$126.
However, due to the limit of proper projectiles, the superheavy
elements synthesized are always neutron deficient and lie in the
deformed region. The deformation effects are very important to
understand the nuclear structures in superheavy nuclei
\cite{cwiok.99,meng.00,ren.01}. It is known experimentally that
the heavy nuclei of the actinum series (Z=93-103) are well
deformed and Bohr and Mottelson \cite{bohr.75} also pointed out
that the deformation can increase the stability of the heavy
nuclei. The microscopic and self-consistent relativistic mean
field model due to its natural description of spin-orbit
interaction \cite{MSY98,MSY99,MT99}, which is a purely
relativistic effect, has been proved to be a reliable method to
describe exotic and superheavy nuclei
\cite{meng.00,long.02,ren.01,bend.99,lala.96}.

In the present work, the recently developed deformed RMF+BCS
method with a density-independent delta-function interaction in
the pairing channel \cite{geng.03} is adopted to analyze
properties of lately synthesized superheavy nuclei $^{288}$115,
$^{287}$115 and their $\alpha$-decay daughter nuclei. The
density-independent (or dependent) delta-function interaction has
been proved to be very successful to take into account the
continuum effect both in relativistic and non-relativistic
self-consistent mean field models \cite{geng.03, yadav.02,
sand.03, meng.98, meng98prl,sand.00}. In the mean field part, the
TMA parameter set \cite{sugahara.94} is used, which has been
proved to be very successful in describing superheavy nuclei
\cite{meng.00,long.02,ren.01} in the relativistic mean field
model. Particular emphasis is put on the effective interactions
used in the particle-particle channel and blocking effects. In
what follows, we discuss in detail numerical details and results
of our calculations.

\section{Numerical details and Results}
The RMF calculations have been carried out using the model
Lagrangian density with nonlinear terms both for the $\sigma$ and
$\omega$ mesons as described in detail in Ref. \cite{geng.03,
sugahara.94}, which is given by
\begin{equation}
\begin{array}{lll}
\mathcal{L} &=& \bar \psi (i\gamma^\mu\partial_\mu -M) \psi +
\,\frac{\D 1}{\D 2}\partial_\mu\sigma\partial^\mu\sigma-\frac{\D
1}{\D 2}m_{\sigma}^{2} \sigma^{2}- \frac{\D 1}{ \D
3}g_{2}\sigma^{3}-\frac{\D 1}{\D
4}g_{3}\sigma^{4}-g_{\sigma}\bar\psi
\sigma \psi\\
&&-\frac{\D 1}{\D 4}\Omega_{\mu\nu}\Omega^{\mu\nu}+\frac{\D 1}{\D
2}m_\omega^2\omega_\mu\omega^\mu +\frac{\D 1}{\D
4}g_4(\omega_\mu\omega^\mu)^2-g_{\omega}\bar\psi
\gamma^\mu \psi\omega_\mu\\
 && -\frac{\D 1}{\D 4}{R^a}_{\mu\nu}{R^a}^{\mu\nu} +
 \frac{\D 1}{\D 2}m_{\rho}^{2}
 \rho^a_{\mu}\rho^{a\mu}
     -g_{\rho}\bar\psi\gamma_\mu\tau^a \psi\rho^{\mu a} \\
      && -\frac{\D 1}{\D 4}F_{\mu\nu}F^{\mu\nu} -e \bar\psi
      \gamma_\mu\frac{\D 1-\tau_3}{\D 2}A^\mu
      \psi,\\
\end{array}
\end{equation}
where all symbols have their usual meanings. The corresponding
Dirac equation for nucleons and Klein-Gordon equations for mesons
obtained with the mean field approximation are solved by the
expansion method on the widely-used axially-deformed
Harmonic-Oscillator basis \cite{geng.03, gambhir.90}. The number
of basis used for expansion is chosen as $N_f=N_b=20$. More basis
have been tested for convergence considerations. We use the
parameter set, TMA \cite{sugahara.94}, for the RMF Lagrangian.

Three kinds of approaches to take into account the pairing
correlations have been adopted in the present work. The first is
the usual RMF+BCS calculation with constant pairing interaction.
The inputs of pairing gaps are $\Delta_n=\Delta_p=11.2/\sqrt{A}$
and the blocking effect is ignored. Such calculations have been
performed extensively by Ren et al. \cite{ren.01}. The second is
the RMF+BCS calculation with a density-independent delta-function
interaction, $V=-V_0\delta(\vec{r}_1-\vec{r}_2)$ \cite{geng.03}.
Here, the blocking effect is also ignored for comparison. The
third is the same as the second one except that the blocking
effect is taken into account by the usual blocking method
\cite{geng.03, geng.032, ring.80}. The pairing force strengths
$V_0$ are fixed by obtaining similar binding energy for
$^{288}$115 as the first approach, i.e., $V_0=280$ MeV fm$^3$ in
the second and $V_0=330$ MeV fm$^{3}$ in the third calculations
respectively. The same $V_0$ has been used for both protons and
neutrons. A slight change of the pairing strength, say ten
percent, only changes the absolute binding energy less than 1.0
MeV and other results are hardly changed. Throughout the paper,
the fist, second, and third kind of calculations will be
abbreviated by "Const", "Delta1" and "Delta2".



\subsection{$\alpha$-decay energies}

In Table. \ref{table1}, a comparison for binding energies and
$\alpha$-decay energies between our three calculations, Const,
Delta1, Delta2, the results from the macroscopic-microscopic
FRDM+FY model \cite{moller.97} and the experimental values for the
$^{288}115$ chain and the $^{287}115$ chain is tabulated. More
detailed properties obtained from the calculations Delta2 are
shown in Table. \ref{table2}. 
The theoretical half-lives $T_\alpha$ are calculated with the
well-known Viola-Seaborg formula \cite{viola.66,moller.97}. The
difference between the predicted $Q_\alpha$ by Const, Delta1,
Delta2, the FRDM+FY model \cite{moller.97}, the YPE+WS model
\cite{munt.03} and the experimental value, $ \Delta_\alpha\mbox
{(theo.)}=Q_\alpha\mbox{(theo.)}-Q_\alpha\mbox{(expt.)}$, is
plotted in Fig. 1 and Fig. 2 for the $^{288}$115 chain and the
$^{287}$115 chain , respectively.

First, for the $^{288}$115 chain, we notice that the quality of
agreement between our calculations ( particularly Delta2 ) and the
experimental values is similar to those of MM models (FRDM+FY and
YPE+WS). For the last two nuclei in the $^{288}115$ chain,
$^{272}$107 and $^{276}$109, results of MM models are closer to
experimental values. For $^{280}111$, our calculations are between
the FRDM+FY model and the YPE+WS model. For $^{284}$113, predicted
$\alpha$-decay energy by our calculations, similar to that of the
YPE+WS model, is larger than the experimental value while the
FRDM+FY model predicts a smaller value. The biggest difference,
about 1.0 MeV, between theory and experiment is also found for
this nucleus. For $^{288}$115, both our calculations and the
FRDM+FY model predict similar values that are smaller than the
experimental value, while the result from the YPE+WS model is
larger than the experimental value.

Second, for the $^{287}$115 chain, similar things happen. For
$^{271}107$, Delta2 and the FRDM+FY model predict similar
$Q_\alpha$, while the YPE+WS model predicts a larger value.
Because no experimental value is observed for this nucleus,
prediction of Delta2 is taken as the experimental value for
comparison. For $^{275}109$, predictions of all our three
calculations deviate from the experimental value more than those
of the MM models. While for $^{279}$111 and $^{287}$115, our
calculations are closer to experimental values than the MM models.
For $^{283}$113, just like the case of $^{284}$114, the difference
between theory and experiment is relatively large. Our
calculations and the YPE+WS model predict different trend for this
nucleus from the FRDM+FY model also.

Third, we note that all our three calculations predict similar
$\alpha$-decay energies for both the $^{288}$115 chain and the
$^{287}$115 chain. The calculations Const and Delta1 give
essentially the same results for both decay chains. While the
calculations Delta2 are generally better than the other two
calculations. This is more obvious for the odd-odd $^{288}$115
chain than for the odd-even $^{287}$115 chain. Since the main
difference between the second and the third calculations is the
blocking effect, we could safely conclude that a proper blocking
treatment can improve the calculated observables for odd-even or
odd-odd nuclei. We also can say that the RMF+BCS calculations with
constant pairing can describe even-even superheavy nuclei
reasonably well. For odd-odd and odd-even nuclei, the inclusion of
the blocking effect becomes necessary. We will see this point more
clearly in the following discussions.

\begin{figure}[t]
\centering
\includegraphics[scale=0.8]{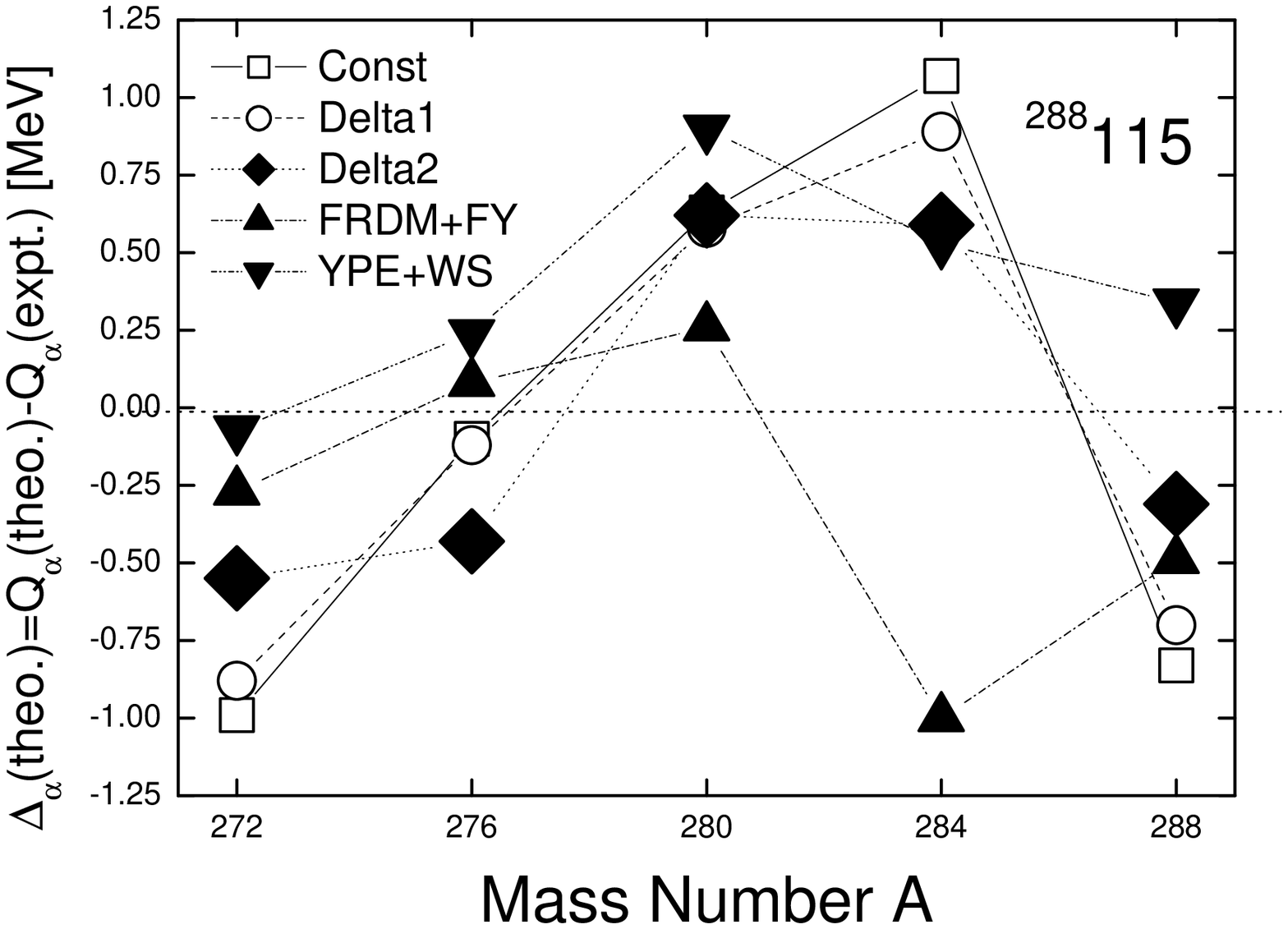}
\caption{\label{fig1.fig}The difference between calculated
$Q_\alpha$(theo.) and experimental $Q_\alpha$(expt.),
$\Delta_\alpha\mbox{(theo.)}=Q_\alpha(theo.)-Q_\alpha(expt.)$, for
the $^{288}115$ chain is plotted against mass number A.}
\end{figure}

\begin{figure}[t]
\centering
\includegraphics[scale=0.8]{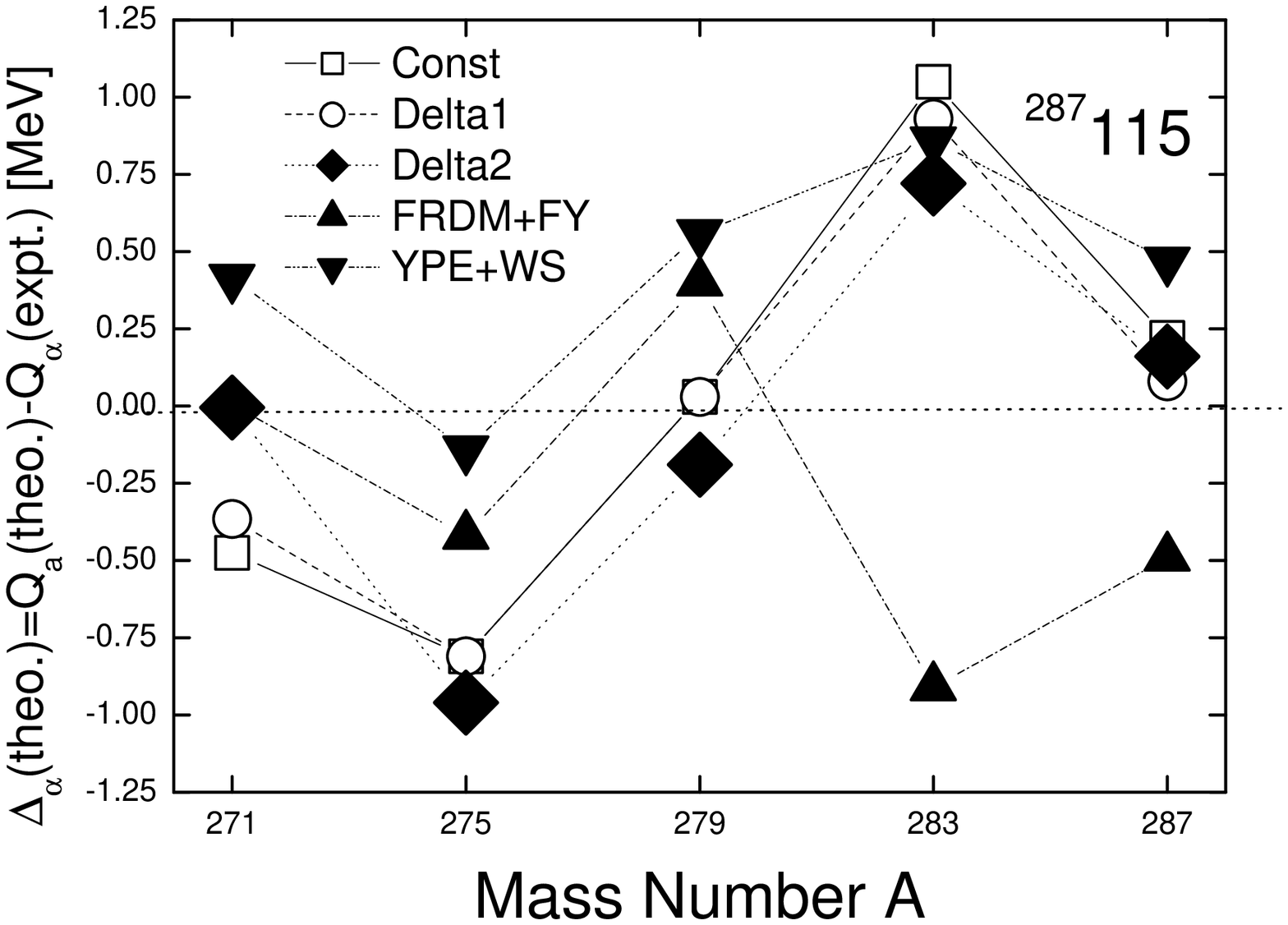}
\caption{\label{fig2.fig}The difference between calculated
$Q_\alpha$(theo.) and experimental $Q_\alpha$(expt.),
$\Delta_\alpha\mbox{(theo.)}=Q_\alpha(theo.)-Q_\alpha(expt.)$, for
the $^{287}115$ chain is plotted against mass number A.}
\end{figure}

Last, we would like to point out the difference between the MM
models and the relativistic mean field theory used here. As we
have seen in the above discussions (see also Table. \ref{table1}
and Fig. \ref{fig1.fig}-\ref{fig2.fig}), predictions of the MM
models are closer to the experimental values for the elements 109
and 107, while for the other three elements, our calculations are
better than those of the MM models. The reason could be that the
MM models depend more on the knowledge of known nuclei. While the
RMF model, due to its natural description of spin-orbit
interaction, after including deformation, pairing interaction and
blocking effect properly, could be more powerful in predicting the
properties of unknown nuclei.

\subsection{Energy surfaces and ground-state deformations}

Now let us discuss a bit more about the differences between our
three different kinds of calculations. We have performed the
constrained quadrupole calculations \cite{geng.03,flocard.73} for
both the $^{288}$115 chain and the $^{287}$115 chain in all the
three calculations. The corresponding energy curves are shown in
Fig. 3 and Fig. 4. We should mention that such calculations are
very time-consuming. First thing we see is that Const and Delta1
give quite similar energy curves for both decay chains. In fact
they also give almost the same results for all calculated
quantities (see also Table. \ref{table1} and Fig. 1-2), except for
the $\alpha$-decay energies where Delta1 is better. Another
noticeable difference is that the hight of the barrier between
different minima can differ a little bit. In most cases, Delta1
give shallower barriers than Const. Second we could see the
difference between calculations with and without blocking, Delta2
and Delta1, is relatively large, despite that the ground-state
properties are quite similar. This shows that proper treatment of
blocking effect is definitely necessary for odd-odd or odd-even
nuclei (see also Fig. \ref{fig1.fig}-\ref{fig2.fig}). For the
$^{287}$115 chain, due to the way that we fixed the pairing
strength $V_0$, the absolute binding energies from calculations
with and without blocking differ around 1.0 MeV for some nuclei.

\begin{figure}[t]
\centering
\includegraphics[scale=0.8]{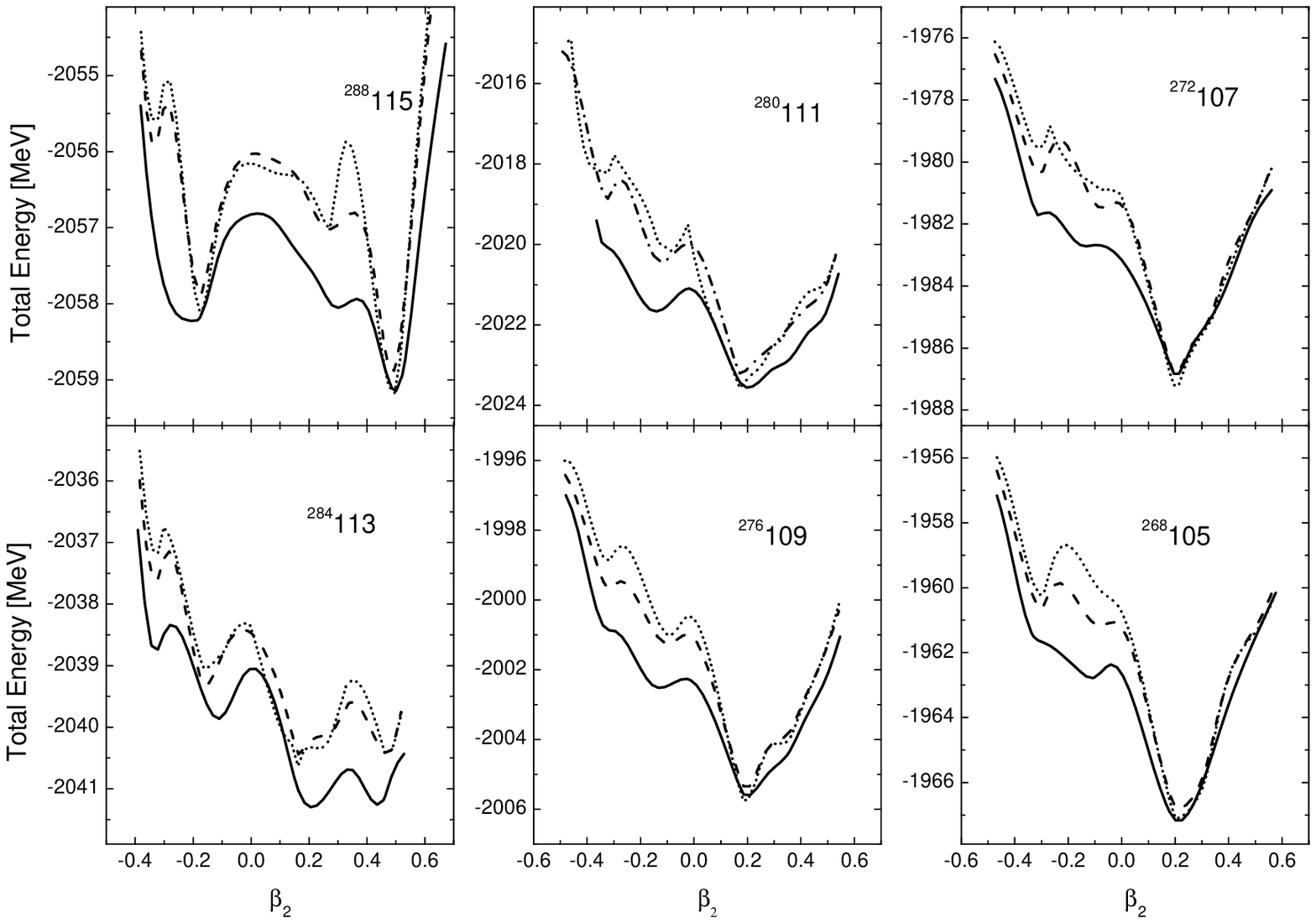}
\caption{\label{fig3.fig}The energy surfaces for $^{288}115$,
$^{284}113$, $^{280}111$, $^{276}109$, $^{272}107$ and $^{268}105$
are plotted as a function of mass quadrupole deformation,
$\beta_2$, for three calculations: Delta2 (solid line), Delta1
(dashed line) and Const (dotted line).}
\end{figure}

\begin{figure}[t]
\centering
\includegraphics[scale=0.8]{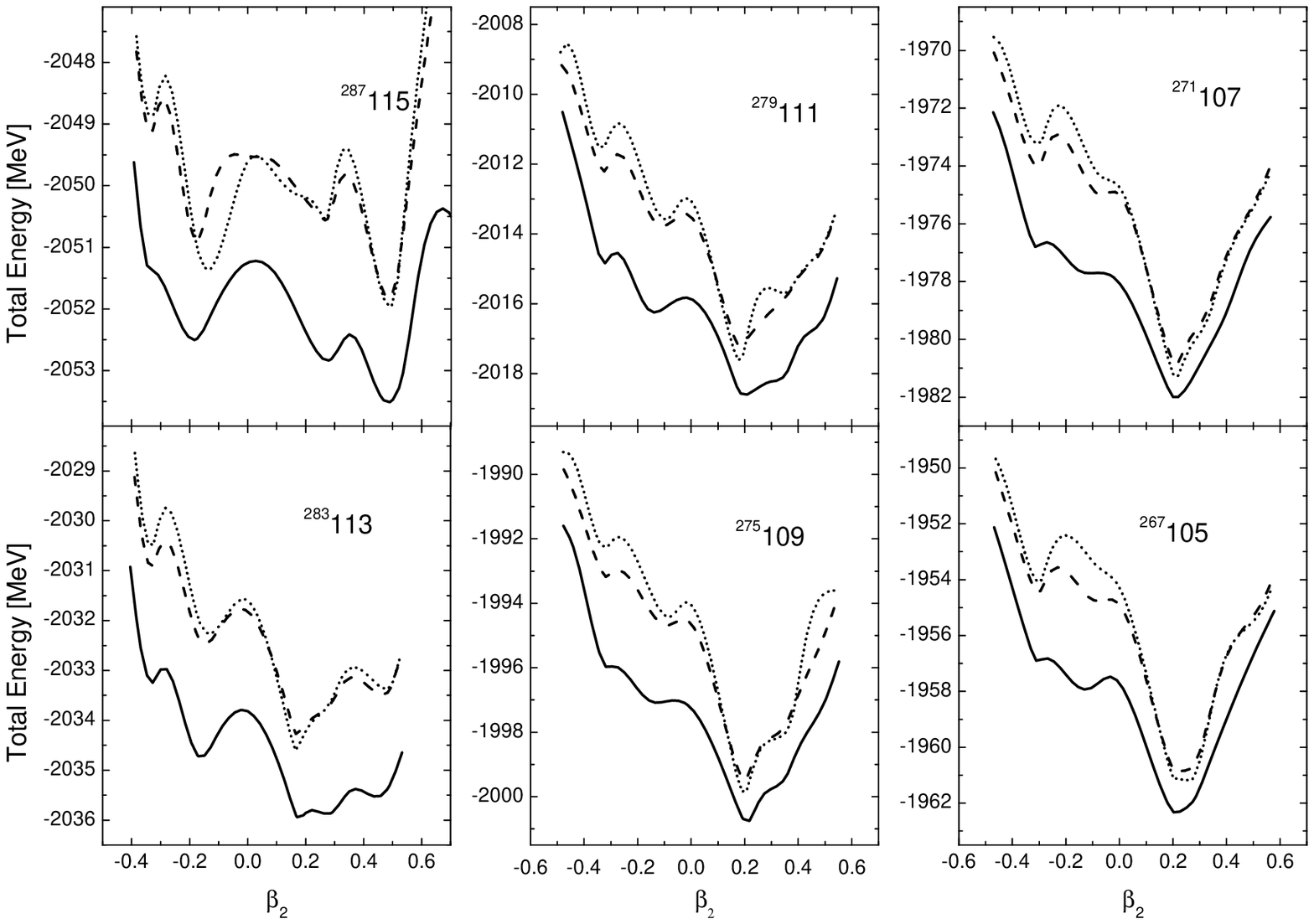}
\caption{\label{fig4.fig}The energy surfaces for $^{287}115$,
$^{283}113$, $^{279}111$, $^{275}109$, $^{271}107$ and $^{267}105$
are plotted as a function of mass quadrupole deformation,
$\beta_2$, for three calculations: Delta2 (solid line), Delta1
(dashed line) and Const (dotted line).}
\end{figure}

\begin{figure}[t]
\centering
\includegraphics[scale=0.8]{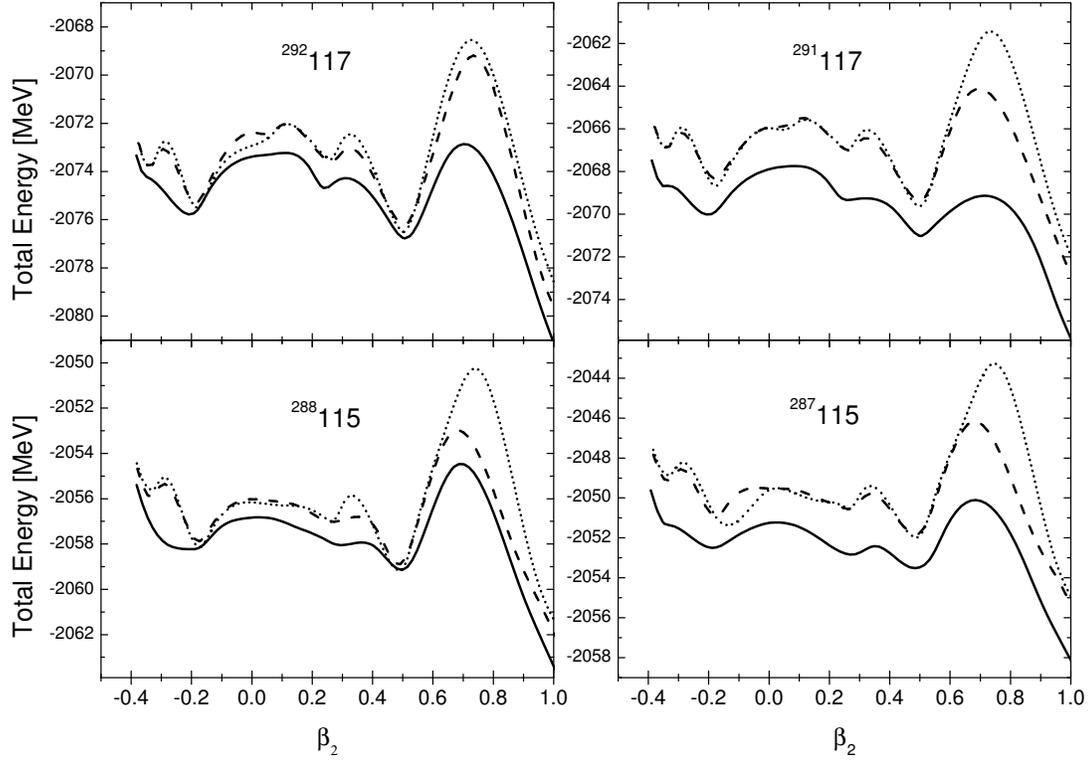}
\caption{\label{fig5.fig}The energy surfaces for $^{292}117$,
$^{288}115$, $^{291}117$, and $^{287}115$ are plotted as a
function of mass quadrupole deformation, $\beta_2$, for three
calculations: Delta2 (solid line), Delta1 (dashed line) and Const
(dotted line).}
\end{figure}

Unlike medium or light nuclei where generally only two minima (one
oblate minimum and one prolate minimum) or one spherical minimum
are observed, the energy curves of superheavy nuclei are more
complicated as we can see in Fig. \ref{fig3.fig}-\ref{fig5.fig}.
This is not surprising. As there are more levels in heavy nuclei,
level crossing is more frequent to happen and lots of local minima
may appear. Except for $^{288}$115 and $^{284}$113 in the
$^{288}$115 chain, $^{287}$115 and $^{283}$113 in the $^{287}$115
chain, the ground state of other nuclei can be determined without
ambiguity, i.e., around $\beta_2\sim0.2$. Similar results have
been obtained by the macroscopic-microscopic YPE+WS model
\cite{munt.03}, more specifically, $\beta_2=$0.200, 0.211 and
0.224 for $^{280}$111, $^{276}$109 and $^{272}$107;
$\beta_2=$0.202, 0.215 and 0.228 for $^{279}$111, $^{275}$109 and
$^{271}$107. The YPE+WS model predicts $\beta_2=0.138$ and
$\beta_2=0.149$ for $^{284}$113 and $^{283}113$, which are also
close to our calculations $\beta_2=0.17$ and $\beta_2=0.18$. While
for $^{288}$115 and $^{287}$115, the YPE+WS model predicts
$\beta_2=0.072$ and $\beta_2=0.066$ respectively, which are quite
different from our calculations, $\beta_2\sim0.5$ for both these
two nuclei. This difference can be understood easily because these
MM models predict $^{298}_{184}114$ to be the next spherical
doubly magic nucleus, while most self-consistent models shift this
property to the more proton-rich side \cite{bend.99}. Further
experiments are needed to clarify these discrepancies between
different theoretical models and different parameter sets in the
same model. In our calculations, two other configurations
$\beta_2\sim-0.2$ and $\beta_2\sim0.3$ are also possible for
$^{288}115$ under certain conditions such as the case of
$\alpha$-decay. That is to say, decay from these two
configurations to $^{283}$113 are also possible based on the
calculated $\alpha$-decay energy. For $^{283}$113, we find that
the minima around $\beta_2\sim 0.2$ and $\beta_2\sim 0.5$ are
close to each other.

Since we see that isotopes of the element 115 are very deformed in
our calculations, we would like to have a closer look at this
element and the element 117, the mother element of the element 115
in the $\alpha$-decay chain. The corresponding energy surfaces
from all the three calculations are plotted in Fig. \ref{fig5.fig}
for $^{292}117$, $^{288}115$, $^{291}117$ and $^{287}115$. It is
clearly seen that the configuration around $\beta_2\sim 0.5$ is
still stable against fission even for the element 117, but the
barrier is lowered greatly for the calculation Delta2 than the
other two calculations. Such an influence to the fission barrier
introduced by the blocking effect has been demonstrated by Rutz et
al. \cite{rutz.99} in the RMF model. Here, we notice that the
adoption of the density-independent delta-function interaction
instead of the constant pairing in the pairing channel further
reduces the fission barrier. Further calculations by Delta2 show
that $\alpha$-decay energies of $^{292}117$ and $^{291}117$ are,
respectively, 10.71 MeV ($B$=2076.62 MeV) and 10.83 MeV
($B$=2053.36 MeV), with $T_\alpha=2.23$ s and $T_\alpha=0.49$ s.
It would be very interesting to synthesize the nuclei $^{292,
291}117$ and measure the $\alpha$-decay chains, since our
calculations predict that these nuclei would make $\alpha$ decays.

To summarize, the constrained calculations show that the energy
curves are relatively complicated and the predicted ground-state
deformations by our calculations are close to those by the YPE+WS
model. Further comparisons show that the fission barriers are
quite different for our three calculations, especially for
calculations with and without blocking. This suggests that to
study superheavy nuclei more reliably one needs to use a more
realistic effective interaction in the pairing channel other than
the const pairing interaction and at the same time include the
blocking effect properly.
\section{Conclusion}
With the recently developed deformed RMF+BCS method with a
density-independent delta-function interaction in the pairing
channel, properties of the lately synthesized superheavy nuclei
$^{288}$115, $^{287}$115 and their $\alpha$-decay daughter nuclei
in Dubna \cite{ogan.03} have been studied. The TMA parameter set
is used for the mean field Lagrangian. In the particle-particle
channel, three different treatments are introduced, i.e., the
usual constant pairing without blocking, the delta-function
interaction with and without blocking. Constrained quadrupole
calculations have been performed also for all these three
calculations. Relatively complicated energy curves are observed
for these superheavy nuclei, especially those nuclei with proton
number larger than 111. The calculated $\alpha$-decay energies,
$Q_\alpha$, are found to agree well with the experimentally
observed values and also are close to those of
macroscopic-microscopic FRDM+FY model and YPE+WS model. Predicted
ground-state deformations agree well with those of
macroscopic-microscopic YPE+WS model. For odd-odd and odd-even
superheavy nuclei, which we have studied here, the inclusion of
the blocking effect can improve the overall performance and is
thought to be necessary based on our calculations. For purposes
other than studying the ground-state properties, a more realistic
interaction in the pairing channel, such as the
density-independent delta-function interaction used in the present
work, would be necessary.

\section{Acknowledgments}

L.S. Geng is grateful to the Monkasho fellowship for supporting
his stay at Research Center for Nuclear Physics where this work is
done . This work was partly supported by the Major State Basic
Research Development Program Under Contract Number G2000077407 in
China and the National Natural Science Foundation of China under
Grant No. 10025522, 10221003 and 10047001.

\newpage
\begin{table}[htbp]
\setlength{\tabcolsep}{0.6 em} \caption{The binding energies, $B$,
and $\alpha$-decay energies, $Q_\alpha$, of decay chains of
$^{288}$115 and $^{287}$115. Listed are the RMF+BCS calculations
with constant pairing, Const, with the delta-function interaction
without blocking, Delta1, and with Blocking, Delta2. FRDM+FY are
results from the finite-range droplet model with folded Yukawa
single-particle potentials \cite{moller.97}. The last column is
the experimental $Q_\alpha$ from Dubna \cite{ogan.03}. All
energies are in units of MeV.}
\begin{center}\label{table1}
\begin{tabular}{c@{\hspace{2ex}}|cc|cc|cc|cc|c}
\hline\hline
Nuclei&\multicolumn{2}{c|}{Const}&\multicolumn{2}{c|}{Delta1}&\multicolumn{2}{c|}{Delta2}&\multicolumn{2}{c|}{FRDM+FY}&Experiment\\
\hline
 &$B$&$Q_\alpha$&$B$&$Q_\alpha$&$B$&$Q_\alpha$&$B$&$Q_\alpha$&$Q_\alpha$\\
 \hline
$^{288}115$&2059.10&9.78&2058.80&9.91&2059.03&10.30&2059.12&10.12&$10.61\pm0.06$\\
$^{284}113$&2040.58&11.22&2040.41&11.04&2041.03&10.74&2040.95&9.15&$10.15\pm0.06$\\
$^{280}111$&2023.50&10.50&2023.15&10.45&2023.47&10.49&2021.81&10.13&$9.87\pm0.06$\\
$^{276}109$&2005.70&9.75&2005.30&9.73&2005.66&9.42&2003.64&9.93&$9.85\pm0.06$\\
$^{272}107$&1987.15&8.16&1986.73&8.27&1986.78&8.60&1985.27&8.88&$9.15\pm0.06$\\
$^{268}105$&1967.01&&1966.70&&1967.08&&1965.86&&\\
\hline
$^{287}115$&2051.88&10.96&2051.72&10.82&2053.36&10.90&2052.72&10.25&$10.74\pm0.09$\\
$^{283}113$&2034.54&11.31&2034.24&11.19&2035.96&10.98&2034.68&9.35&$10.26\pm0.09$\\
$^{279}111$&2017.55&10.55&2017.13&10.55&2018.64&10.33&2015.73&10.92&$10.52\pm0.16$\\
$^{275}109$&1999.80&9.67&1999.38&9.67&2000.67&9.52&1998.36&10.06&$10.48\pm0.09$\\
$^{271}107$&1981.17&8.18&1980.75&8.29&1981.89&8.65&1980.13&8.66&\\
$^{267}105$&1961.05&&1960.74&&1962.24&&1960.49&&\\
\hline\hline
\end{tabular}
\end{center}
\end{table}
\newpage
\begin{table}[htbp]
\setlength{\tabcolsep}{0.6 em} \caption{The binding energies, $B$,
neutron and proton quadrupole deformations, $\beta_{2n}$ and
$\beta_{2p}$, neutron and proton rms radii, $R_n$ and $R_p$, the
calculated alpha-decay energies and life-lives, $Q_\alpha$ and
$T_\alpha$, of superheavy nuclei on the alpha-decay chains of
$^{288}$115 and $^{287}$115 from the calculations Delta2. The last
two columns are experimental decay energies and liftimes. All
energies are in units of MeV and all radii in units of Fermi.}
\begin{center}\label{table2}
\begin{tabular}{c@{\hspace{2ex}}|@{\hspace{2ex}}c@{\hspace{2ex}}|@{\hspace{2ex}}cccc@{\hspace{2ex}}|@{\hspace{2ex}}cc@{\hspace{2ex}}|@{\hspace{2ex}}cc}
\hline\hline
Nuclei&$B$&$\beta_{2n}$&$\beta_{2p}$&$R_n$&$R_p$&$Q_\alpha$&$T_\alpha$&$Q_\alpha$(expt)&$T_\alpha$(expt)\\
\hline
$^{288}115$&2059.03&0.48&0.50&6.58&6.41&10.30&6.86 s&$10.61\pm0.06$&$87^{+105}_{-30}$ ms \\
$^{284}113$&2041.03&0.17&0.17&6.37&6.18&10.74&111.96 ms&$10.15\pm0.06$&$0.48^{+0.58}_{-0.17}$ s \\
$^{280}111$&2023.47&0.18&0.19&6.34&6.15&10.49&118.97 ms&$9.87\pm0.06$&$3.6^{+4.3}_{-1.3}$ s \\
$^{276}109$&2005.66&0.20&0.20&6.32&6.12&9.42&25.08 s&$9.85\pm0.06$ &$0.72^{+0.87}_{-0.25}$ s \\
$^{272}107$&1986.78&0.20&0.21&6.30&6.09&8.60&1953.31 s&$9.15\pm0.06$ &$9.8^{+11.7}_{-3.5}$ s \\
$^{268}105$&1967.08&0.21&0.22&6.28&6.06&&& &$16^{+19}_{-6}$ h \\
\hline
$^{287}115$&2053.36&0.48&0.50&6.56&6.41&10.90&80.57 ms&$10.74\pm0.09$&$32^{+155}_{-14}$ ms \\
$^{283}113$&2035.96&0.18&0.18&6.36&6.18&10.98&12.76 ms&$10.26\pm0.09$&$100^{+490}_{-45}$ ms \\
$^{279}111$&2018.64&0.20&0.20&6.34&6.15&10.33&142.07 ms&$10.52\pm0.16$&$170^{+810}_{-80}$ ms \\
$^{275}109$&2000.67&0.21&0.21&6.32&6.12&9.52&5.77 s&$10.48\pm0.09$ &$9.7^{+46.}_{-4.4}$ s \\
$^{271}107$&1981.89&0.21&0.21&6.29&6.09&8.65&604.91 s&& \\
$^{267}105$&1962.24&0.22&0.22&6.27&6.06&&& &$73^{+350}_{-33}$ min \\
 \hline\hline
\end{tabular}
\end{center}
\end{table}

\newpage


\end{document}